# Dynamic metasurface lens based on MEMS Technology

Tapashree Roy[1,#,*], Shuyan Zhang[2,*], IL Woong Jung[1], Mariano Troccoli[3], Federico Capasso[2], and Daniel Lopez[1]

(*Equal Contributors)

[1]Nanoscience and Technology Division, Argonne National Laboratory, Lemont, IL 60439

[2]John A. Paulson School of Engineering and Applied Sciences, Harvard University, Cambridge, MA 02138

[3]Evolution Photonics Inc., Pasadena, CA 91106

[#]Present address: Applied Materials Inc., Mumbai 400076, India

## Abstract

In the recent years, metasurfaces, being flat and lightweight, have been designed to replace bulky optical components with various functions. We demonstrate a monolithic Micro-Electro-Mechanical System (MEMS) integrated with a metasurface-based flat lens that focuses light in the mid-infrared spectrum. A two-dimensional scanning MEMS platform controls the angle of the lens along the two orthogonal axes (tip-tilt) by ±9 degrees, thus enabling dynamic beam steering. The device can compensate for off-axis incident light and thus correct for aberrations such as coma. We show that for low angular displacements, the integrated lens-on-MEMS system does not affect the mechanical performance of the MEMS actuators and preserves the focused beam profile as well as the measured full width at half maximum. We envision a new class of flat optical devices with active control provided by the combination of metasurfaces and MEMS for a wide range of applications, such as miniaturized MEMS-based microscope systems, LIDAR scanners, and projection systems.

## 1. INTRODUCTION

Recently, metasurface-based flat optical devices capable of shaping the wavefront of light have come to the forefront of ongoing scientific research [1, 2, 3]. Planar counterparts of conventional bulky optical devices like lenses [4-6], beam deflectors [7, 8], holograms [9], polarimeters [10], and so on have been experimentally demonstrated. These devices use sub-wavelength dimension metallic and/or dielectric phase shifting optical elements arranged on a two-dimensional plane, "metasurface", mimicking the phase profile of the conventional bulk optical device. Such metasurface-based flat devices represent a new class of optical components that are compact and lightweight. However, most of these nanostructured devices are static, which limits the functions that can be achieved.

In this paper, we introduce the concept of dynamically controlling these metasurfaces by integrating them onto Micro-Electro-Mechanical Systems (MEMS). These MEMS have the unique advantages of high-speed movement, excellent optical quality, wavelength and polarization independence, and low optical loss [11]. We present a prototype consisting of a MEMS-controlled reflective metasurface lens that focuses in the mid-infrared spectrum. The preferred MEMS device is an electrostatically controlled 2D scanner micro-mirror because it is a key element used in many applications such as LIDAR laser scanners [12], optical communications [13-14], and bio-imaging [15-18]. Technologies like these will greatly benefit from a dynamically controlled metasurface-based lens because it will facilitate the removal of bulky optical components in the system while allowing unique functions such as the predetermined shaping of light beams. When electrostatically actuated, the MEMS platform controls the angle of the lens along two orthogonal axes allowing scanning of the flat lens focal spot by about ±9 degrees in each direction.

## 2. SAMPLE DESIGN AND FABRICATION

We design a plasmonic lens producing a line focus, like a cylindrical lens, when illuminated with monochromatic mid-infrared light of wavelength $\lambda = 4.6$ µm. As the design unit cells, we choose polarization independent, sub-wavelength-sized, 50 nm thick gold resonators in the shape of a disc (Fig. 1a). The resonators sit on a 400 nm thick silicon dioxide layer deposited on a 200 nm thick continuous gold film. This structure has been used to improve the focusing efficiency in reflection [19, 20]. By changing the radius of the disc, the phase of the reflected light changes (Fig. 1b). To construct a planar lens, we spatially distribute the discs with varying radii to realize the hyperbolic phase profile. Figure 1c shows a schematic of the reflective metasurface lens when a collimated Gaussian beam is incident at an angle θ on the lens and is focused at a distance $f$ along the normal axis to the lens surface. This configuration proves the flexibility of our design technique: by using the metasurface itself to spatially separate the incident and reflected beams, we avoid the need for a beam splitter, resulting already in a reduction in the number of bulk optical components that are conventionally used in optical systems.

The equation of the phase profile used to design the metasurface lens is

$$\varphi(x) = \frac{2\pi}{\lambda}\left(f - \sqrt{x^2 + f^2} - x \cdot \sin\theta\right)$$

where $\lambda$ is the wavelength, $f$ is the focal length and $x$ is the position of the phase shifting elements in the lens. The lens functions as a cylindrical lens, i.e. one-dimensional focusing, so the phase profile is only a function of $x$ [19]. One improvement from previous literature [21] is that the overall packing density of the subwavelength resonators is increased by arranging them with constant edge-edge distance, which helps to improve the sampling of the ideal phase profile. The metasurface lens demonstrated here is designed on a square layout with each side measuring $W = 0.8$ mm to focus light incident at an angle of $\theta = 45°$ to the lens surface at a distance $f = 5$ mm away.

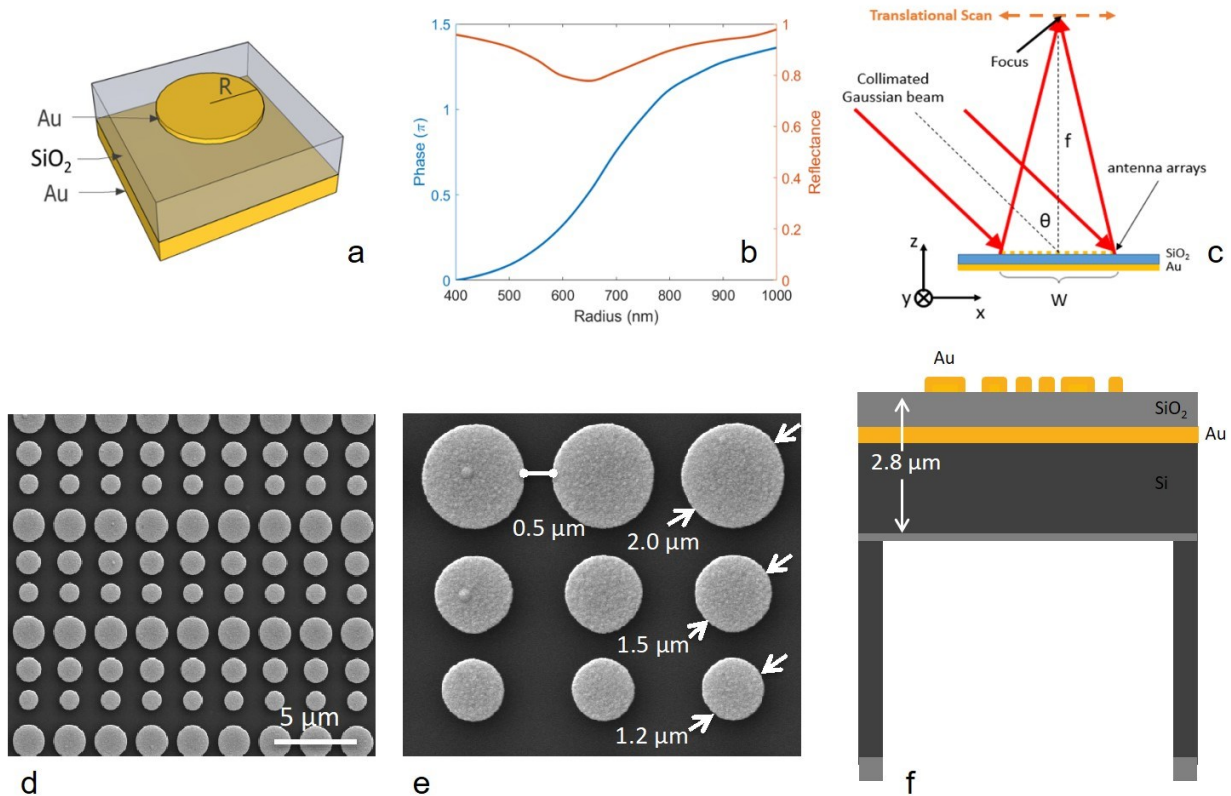

*Figure 1: Design and fabrication of the metasurface lens: (a) A unit cell consisting of a 50 nm thick gold disc on 400 nm thick silicon dioxide substrate with a 200 nm thick gold backplane. (b) Simulated values of reflectance and phase for varying sizes of gold discs. (c) Schematic representation of focusing characteristics of the reflective metasurface lens for tilted illumination. (d, e) Scanning electron microscope images of the fabricated lens. (f) Schematic cross-section of the different constituting layers of the membrane-supported flat lens.*

The metasurface lens is fabricated using standard photolithography techniques on a silicon-on-insulator (SOI) wafer with a 2 μm thick top device layer, a 200 nm buried-oxide layer, and a 600 μm thick handle layer. SEM images of the reflective lens are shown in Figure 1d and 1e. The detailed process flow for fabricating the lens is described in Supplementary Information S1. Following the lens fabrication, the handle layer is removed using a dry-etch process based on xenon-difluoride, resulting in a reflective flat lens standing on top of a 2.8 μm thick membrane (Fig. 1f). For the next stage of fabrication, we use a focused ion beam (FIB) tool integrated with an Omniprobe micromanipulator needle to integrate the flat lens onto a 2D MEMS scanner. Figure 2a depicts the steps of the process. We trace the focused ions around the periphery of the lens to cut out most of the structure, except a small portion. Here we weld the needle-tip of the micromanipulator by depositing platinum. The remaining structure is subsequently released; the membrane is free from the surrounding solid substrate and is held only by the micromanipulator needle. We move the membrane out of the substrate and on to the MEMS platform by controlling the micromanipulator arm. After the lens is placed and aligned with the central platform of the MEMS, it is *glued* by depositing platinum in small patches. Finally the needle is cut away by milling with focused ion beam [22]. This practical technique allows for fusion welding of the flat lens to the MEMS device without the need for extra materials. Moreover, it enables the integration of hybrid structures fabricated with processes having different critical dimensions and structural materials (see Supplementary Information S2).

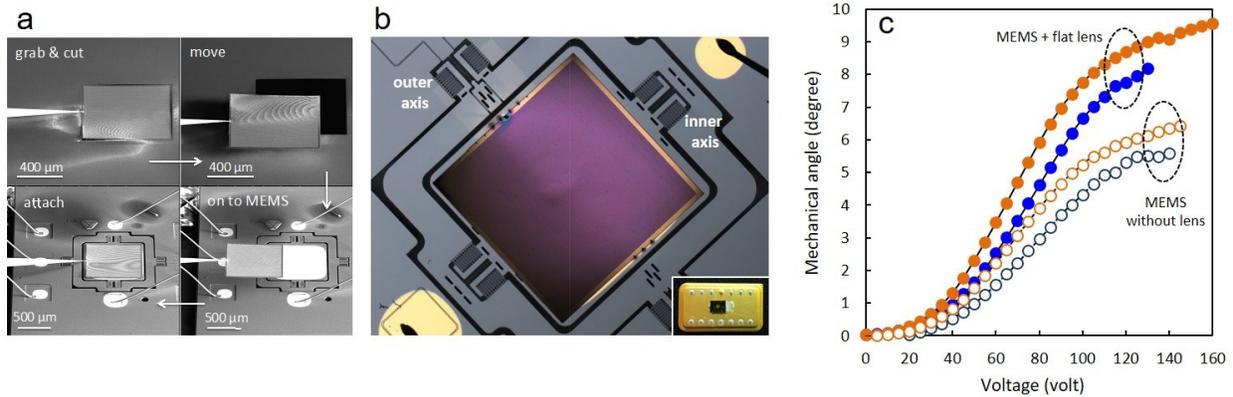

*Figure 2. Integration of the flat lens onto MEMS: (a) Stages of integration of the metasurface-based flat lens with an external 2D MEMS platform. (b) Optical microscope image of a MEMS scanner with a flat lens on top. The two rotational axes of the scanner are indicated. The inset shows the device mounted on a dual in-line package ready for electrostatic actuation. (c) Angular displacement of the MEMS scanner with and without the metasurface-based lens. The orange circles represent the angular displacement when the inner axis is actuated, while the blue circles show the response when the outer axis is activated.*

An optical image of the 2D MEMS scanner with the integrated flat lens is shown in Figure 2b. The scanner is gimbaled for achieving biaxial degrees-of-freedom and is actuated by electrostatic vertical comb-drives. Simple single-layer straight beams are used for torsional flexures on both inner and outer axes to give a 2-D rotational degree-of-freedom. The micromirror is electrostatically rotated about the inner axis using the vertical comb drives mounted on the gimbal frame [15]. The gimbal frame rotates about the outer axis using the vertical comb drives mounted on its frame and substrate. The mirror dimensions are 1 mm x 1mm with a thickness of 10 µm. Although thicker device layers are desired for better flatness of mirrors and frame under dynamic deformation, they are also more difficult to fabricate due to the required high aspect-ratio fabrication process and alignment issues. Simulations showed that 10 µm thick layers provide sufficiently large stable deflection range with minimal adverse effect on the performance of the device. To characterize the mechanical response of the flat lens-on-MEMS assembly, the device is mounted on a dual in-line package and electrostatically actuated to independently control either the inner or the outer axes, as shown in Fig. 2b. Details of the MEMS platform and voltage connections are given in Supplementary Information S3. Figure 2c summarizes the mechanical response of the MEMS scanner when a voltage is applied across each rotational axis. As the applied voltage is increased, the MEMS mirror starts rotating until a saturating region is reached, beyond which the comb drives cannot be moved further. We perform this experiment under an optical profilometer, capturing an image of the MEMS surface for each applied voltage. The rotation angle of the MEMS mirror is calculated with respect to a rigid peripheral structure. The measurements are taken for bare MEMS, i.e. before the metasurface lens has been integrated, and for the same device after it has been loaded with the metasurface lens. As shown in Fig. 2c, even after the addition of the metasurface lens on top of the MEMS, the functional dependence of the MEMS mirror remains the same. The increased angular dependence of the loaded MEMS is a consequence of using vertical comb drives for torsional actuation of the device. The incorporation of the flat lens onto the micro-mirror reduces the static gap between vertical combs resulting in a much efficient actuation and thus, a larger angular deflection.

## 3. OPTICAL CHARACTERIZATION

The focusing characteristic of the flat lens is simulated by using Finite Difference Time Domain (FDTD) method (Lumerical Inc.). Only one row of nano-disc antennas along the x-direction where the phase profile of the cylindrical lens has been imposed is simulated, while Bloch boundary conditions are applied along the y-directions. Figure 3a is the calculated distribution of the electric field intensity ($|E|^2$) near the focal region in the *x-z* plane. The focal length is determined by the *z* value of the highest intensity point, which is indicated by the white dashed line at $z = f = $ 5 mm. The focusing efficiency is estimated to be 83%, which is calculated from the beam intensity at the focal region normalized to the source power. Figure 3b shows the beam profile at the focal region across the dashed line in Fig. 3a. The simulated discrete data points (blue circles) are fitted with a Gaussian curve (red line) to determine the full width at half maximum (FWHM) of the focal beam, which is 22.8 µm. This is only slightly bigger than the diffraction-limited value (21.6 µm), possibly due to the discrete phase approximation of the ideal 0 to $2\pi$ variation.

For the experimental characterization, we use a continuous-wave Fabry-Pérot quantum cascade laser (QCL) (AdTech Optics) emitting at $\lambda = 4.6$ µm. The laser is mounted so that its output beam is s-polarized (electric field of the light is first sent through a pinhole before reaching the detector). Because of the small focused beam size, another dual-lens system is placed before the detector to expand the focused beam by a factor of 2.5 (see Supplementary Information S4). The signal-to-noise ratio is increased by modulating the intensity of the QCL with a small sinusoidal current superimposed on the direct current (Wavelength Electronics QCL1500) and demodulating the detected signal with a lock-in amplifier (AMETEK Advanced Measurement Technology).

The focus is characterized experimentally by translational scans. The data are taken at the center of the focal line, i.e. $y = 0$. The average standard deviation of repeated measurements is 0.2%. The detector is scanned across the focus in 2 µm steps, which is smaller than the pinhole diameter (10 µm). So, the raw data is de-convoluted to retrieve the original beam profiles. Figure 3c shows the measurement results. The FWHM is 26.2 µm, which is close to the simulated value. The difference is possibly due to fabrication errors and the $M^2$ factor of the QCL not being 1 ($M^2 = 1.2$).

The optical focusing performance of the metasurface lens integrated with the MEMS is experimentally characterized using the same set up as described above. The angle of the incident light is at 45° to the unactuated MEMS scanner. The profile of the reflected focal line is measured for three positions of the MEMS platform: actuating voltages of 0 V (unactuated), 40 V, and 60 V are applied across the outer axis such that the lens tilts by 0°, 1°, and 2.5° respectively (Fig. 3d-f). The measured FWHM of the focal lines for the three tilted positions of the lens are shown in Fig. 3g-i respectively. For each position of the lens, the detector is rotated to align with the peak intensity of the reflected light. To measure the beam profile, a translational scan is performed across the rotated focal line position with a 30 µm pinhole in 5 µm steps. For the designed 0° position of the MEMS platform, when compared to the lens on the solid substrate (Fig. 3c), we observe an increase in FWHM. This is attributed to the non-flatness of the lens when released from the solid substrate. However, tilting the MEMS integrated lens up to 2.5° preserves the focused beam profile, as well as the measured FWHM. The simulation results and analysis are further summarized in Supplementary Fig. S5.

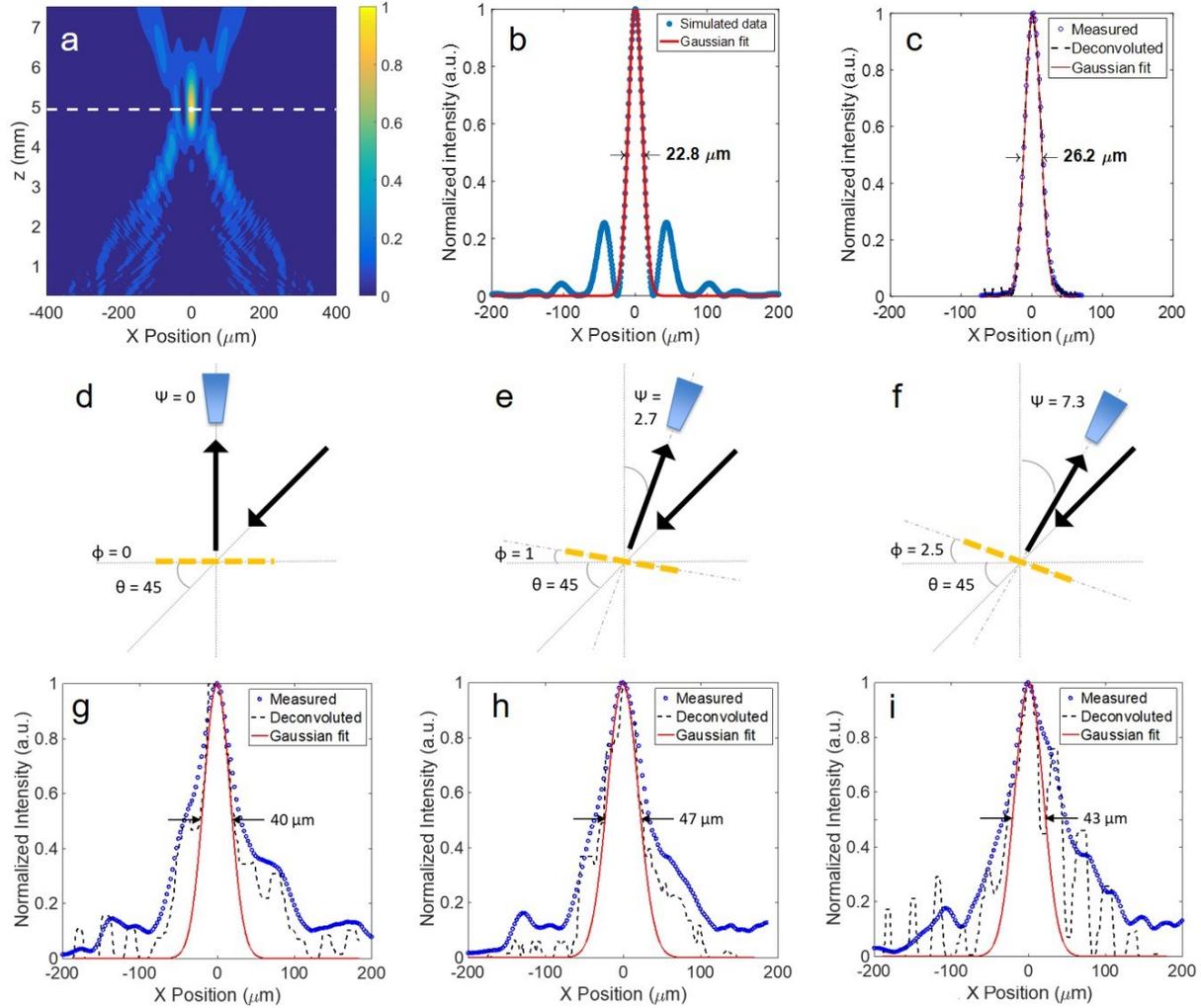

*Figure 3: (a) Simulation: distribution of the intensity (normalized $|E|^2$) of the reflected beam in the x-z plane at y = 0. The lens is centered at x = 0 and the size of the lens is from -0.4 mm to 0.4 mm. The white dashed line indicates the focal length. (b) Simulation: line scan of the focused beam profile at y = 0 and z = f = 5 mm along the white dashed line. (c) Experiment: translational scan of the reflected beam intensity (normalized $|E|^2$) measured at the center of the focal line. (d-f): Schematic of the three experimental configurations: the MEMS scanner is actuated to move the lens by 0°, 1°, and 2.5° respectively while the angle of the incident illumination remains unchanged. To align with the peak of the reflected signal, the position of the detector needs to be at 0°, 2.7°, and 7.3° respectively. (g-i) Optical profile at the focal line of the reflected beam when the actuated lens-on-MEMS device is rotated by 0°, 1°, and 2.5° respectively. Experimentally measured translation scan for each of the three configurations. The FWHMs calculated from the Gaussian fits are comparable, though the asymmetry of the central peak changes with the tilting of the platform.*

In order to quantify the effect of the surface curvature on the otherwise designed *flat* lens performance, we measure the curvature of the lens-on-MEMS assembly using an optical profilometer and simulate the performance of a cylindrical lens on top of a substrate having the same curvature (see Supplementary Information S6 for details). The results are shown in Fig. 4. When supported by the curved substrate, the focal length of the lens increases to 8.9 mm compared to the designed value of 5 mm, and the reflected beam path is at 0.5º compared to the designed value of 0º. The FWHM becomes wider, 41.7 μm compared to 22.8 μm as on a solid substrate, which is in reasonable agreement with the results reported in Figure 3g-i.

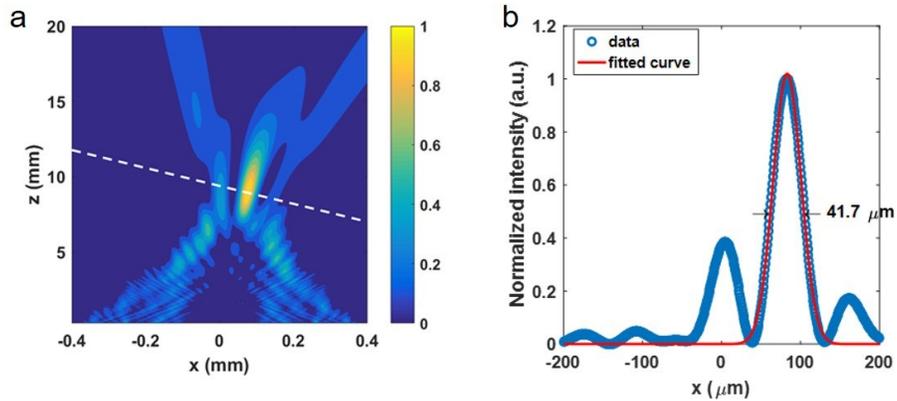

*Figure 4. (a) Distribution of the intensity (normalized $|E|^2$) of the reflected beam in the x-z plane at y = 0 with a curved substrate. The white dashed line is perpendicular to the reflected beam path. (b) Simulated line scan of the focused beam profile along the white dashed line compared with the experimental translational scan of the reflected beam intensity (normalized $|E|^2$) measured at the center of the focal line.*

## 4. CONCLUSION

In summary, we have presented a MEMS-integrated metasurface lens. The device can be electrically controlled to vary the 2D angular rotation of a flat lens and hence the position of the focal spot by several degrees. This proof-of-concept integration of metasurface-based flat lenses with 2D MEMS scanners can be extended to the visible and other parts of the electromagnetic spectrum implying the potential for application across wider fields, such as MEMS-based microscope systems, holographic and projection imaging, LIDAR scanners, laser printing, and so on. Dense integration of thousands of individually controlled lens-on-MEMS devices onto a single silicon chip would lead to the creation of a new type of reconfigurable fast digital SLM [23] that would allow an unprecedented degree of control and manipulation of the optical field.

5. ACKNOWLEDGEMENT

We acknowledge Tobias Mansuripur for the valuable advice on the experiments and Alan She for helpful discussions. We acknowledge the funding support from Air Force Office of Scientific Research (AFOSR) (MURI: FA9550-12-1-0389). Shuyan Zhang acknowledges the funding support from National Science Scholarship from A*STAR, Singapore. This work was performed in part at the Center for Nanoscale Systems (CNS), a member of the National Nanotechnology Infrastructure Network (NNIN), which is supported by the National Science Foundation under NSF award no. ECS-0335765. CNS is part of Harvard University. Use of the Center for Nanoscale Materials was supported by the U. S. Department of Energy, Office of Science, Office of Basic Energy Sciences, under contract No. DE-AC02-06CH11357.

# Supplementary Information

**S1: Lens fabrication**

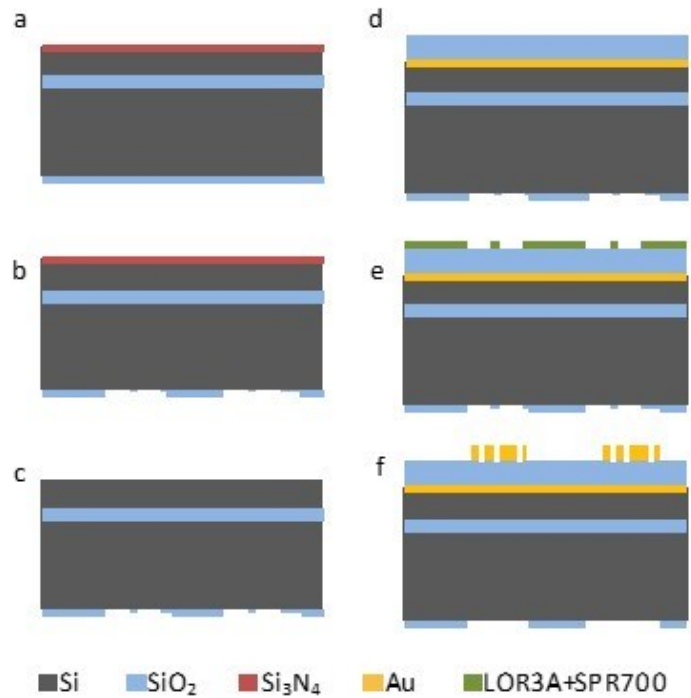

*Figure S1: Fabrication of the planar lens. (a) – (f) Process flow of photolithography as described in the section S1.*

In the following section, we describe the fabrication steps following the figure numbers of Fig. S1:

(a) On the topside of the SOI wafer, a 100 nm thick protective layer of silicon nitride ($Si_3N_4$) is deposited by the plasma-enhanced chemical vapor deposition process. On the backside, a 110 nm thick layer of silicon dioxide ($SiO_2$) is deposited using the same process; this layer would serve as the oxide mask for etching the handle layer at a later stage of the fabrication.

(b) A positive photoresist (SPR 700) is coated on the backside and exposed using a stepper tool (Autostep 200 i-line) to print circular windows, which would be perfectly aligned with the flat lenses to be fabricated on the topside. Using the developed photoresist as an etch-mask, the silicon dioxide layer is plasma-etched to produce the windows.

(c) Next, the topside protective layer of $Si_3N_4$ is cleared in hot phosphoric acid (85% $H_3PO_4$ at 165 °C for 5 minutes).

(d) Now we start fabricating the flat lens on the topside. We deposit 200 nm thick layer of gold (using electron beam vapor deposition), followed by 400 nm thick $SiO_2$ (using PECVD).

(e) We choose a bilayer resist for facilitating clean lift-off. The topside is coated with the photoresists LOR 3A followed by SPR 700. The resist bilayer is exposed using the stepper, ensuring each lens structure is accurately aligned with the previously etched backside windows.

(f) Finally, 50 nm gold with an adhesion layer of 5 nm thick titanium is deposited using e-beam vapor deposition, and lifted-off in remover-PG. This results in the gold disks constituting each lens.

**S2: Integrating the flat lens with MEMS**

To integrate the flat lens with an external MEMS device, we "tear out" the lens from the thick SOI wafer. To do this, we start by etching the 600 μm thick handle layer in xenon difluoride ($XeF_2$). Xenon difluoride etch is a highly selective, isotropic, dry-etch process for silicon [1]. The backside window made with $SiO_2$ provides the entry point for the xenon difluoride gas to react with the bulk of silicon in the handle layer. Due to the excellent selectivity of $XeF_2$ to silicon versus $SiO_2$, the 200 nm thick buried-oxide layer of the SOI wafer also serves as the etch-stop layer. The etching process is visually inspected in-situ until all the silicon directly beneath the lens is etched out, and the lateral extent exceeds the outer dimension of the lens. Figure S2 shows an optical image of the etched portion from the backside as well as from the topside. From the top, the shadow beneath the square lens structure indicates the area where silicon has been etched out. After the $XeF_2$ etch, we are left with a membrane that is only 2.8 μm in thickness and almost 1.35 mm in diameter.

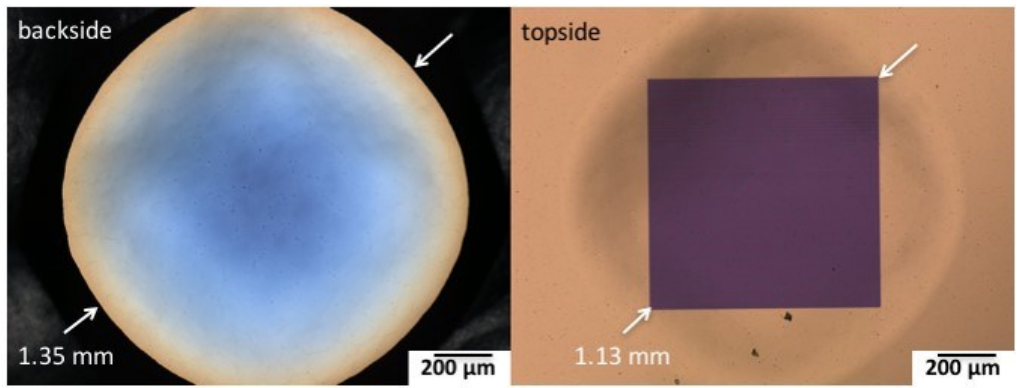

*Figure S2: Optical microscope image after etching of the SOI handle layer from the backside and from the topside. In the topside image, the thinned membrane is a circular shadow surrounding the square metasurface lens.*

## S3: Details of MEMS scanner and voltage connections

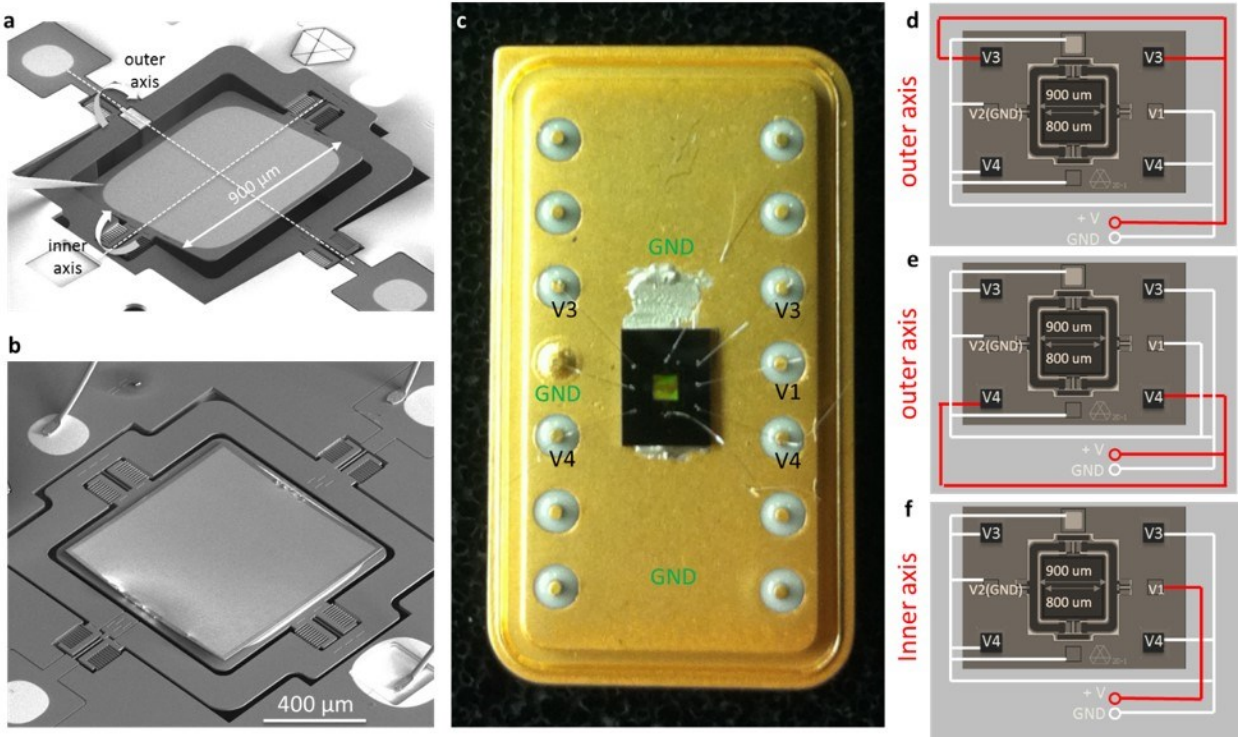

*Figure S3: (a) Scanning electron micrograph (SEM) of a bare MEMS platform with a square layout, each side measuring 900 μm. (b) SEM of a flat lens integrated with the MEMS platform shown in (a). (c) Photograph of the DIP where the MEMS-with-lens is attached with silver paste and wire-bonded appropriately. (d – f) Voltage connections (in red) for actuating outer or inner axis of the MEMS platform.*

## S4: Experimental setup for optical characterization

Figure S4 shows the schematic of the experimental arrangement. The ~3 mm laser beam is downsized by 2.5 times, such that the incident beam is only slightly larger than the 800 μm square lens. The light is incident at 45° to the lens surface. The lens reflects the incident beam at 0° and focuses the light at 5 mm away from its surface. The focused light is magnified 2 times by the lens pair L3 and L4 and the intensity is detected by a thermo-electrically cooled mid-IR detector through a pinhole. The detector, L3, and L4 sit on a rotational stage so that the angle can be adjusted. The detector is mounted on a translational stage to scan the reflected beam.

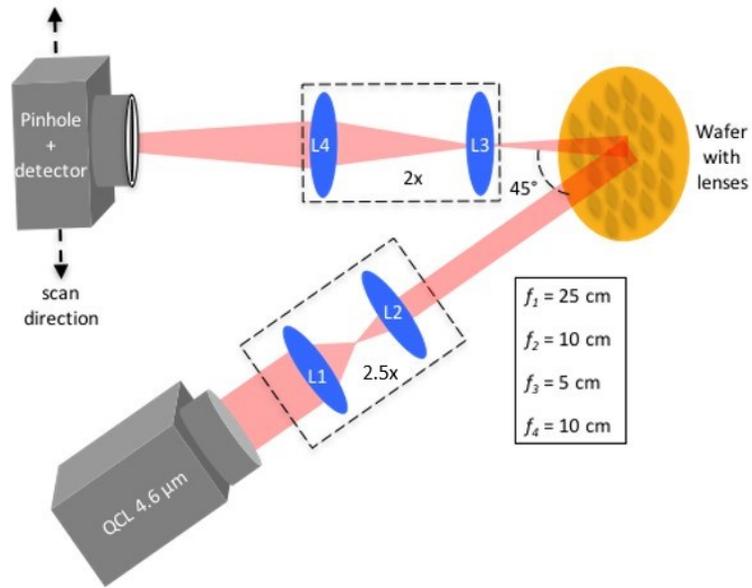

*Figure S4: Schematic showing experimental arrangement used for characterization of the optical response of a flat lens.*

### S5: Simulation of actuated MEMS for off-axis aberration correction

The simulation is performed assuming a flat substrate. The definitions of the various terms used are given in Fig. S5a and the main results are summarized in Fig. S5b.

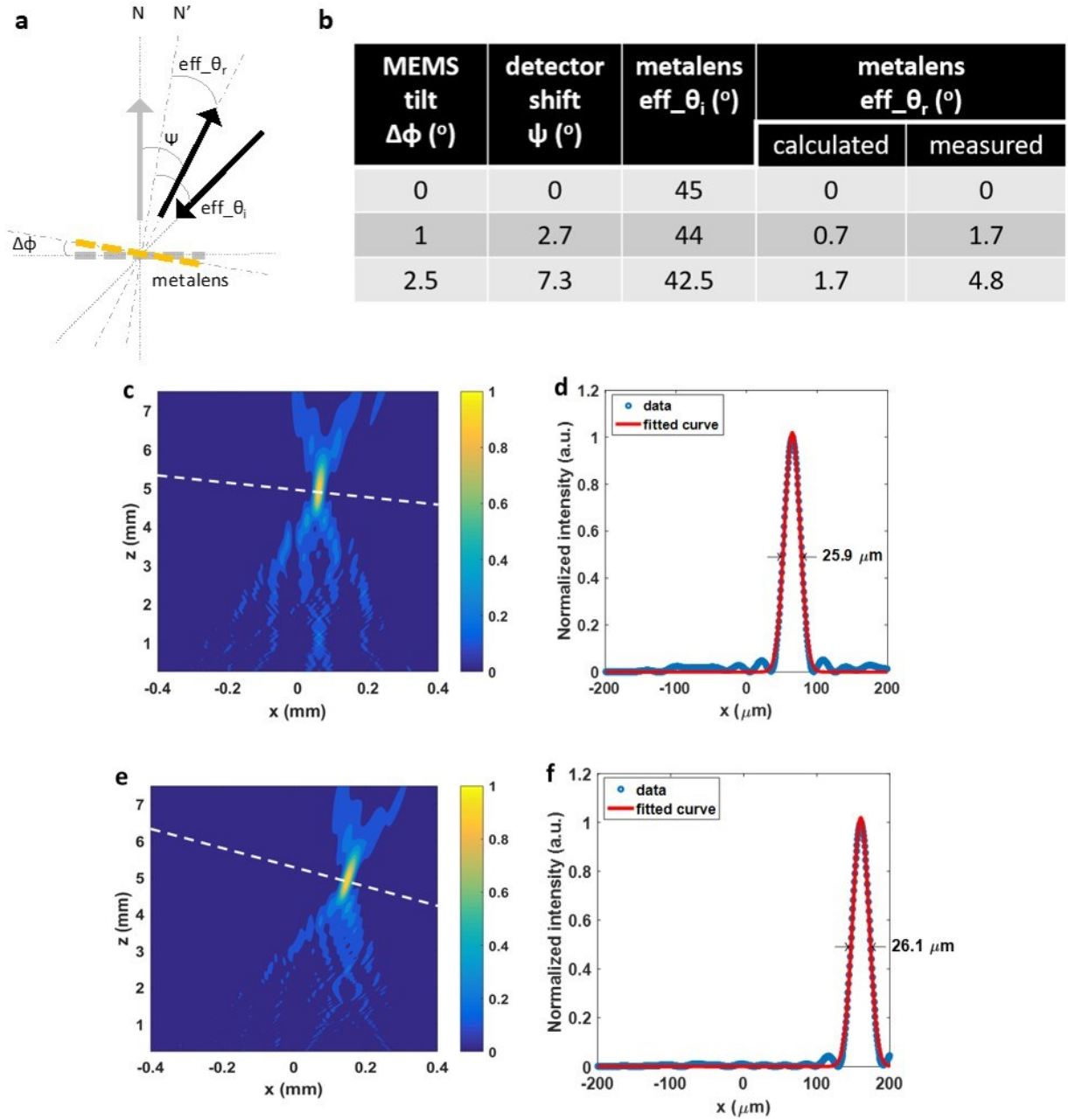

*Figure S5: (a) Schematic for defining the effective incident angle and effective reflective angle, when the lens is tilted. (b) Table comparing calculated (for a flat lens) and measured (for a non-flat lens) effective reflective angles for three different tilt angles of the MEMS platform. (c)-(f) FDTD simulation of reflected beam intensity and the line cut at the focal length for tilt angle: (c, d) 1° and (e, f) 2.5° respectively.*

## S6: Topology measurements of the surface of the integrated device

To analyze the effect of the substrate non-flatness on the focusing performance of the *flat* lens, we used the data of the height profile at the center (see Fig. S6c), along the direction over which the cylindrical phase profile is imposed, and simulated the lens performance on a curved substrate. The results are shown in the main paper Fig. 4. The focal length becomes 8.9 mm compared to the designed value 5 mm. The reflected beam path is 0.5° compared to the designed value 0°. The FWHM becomes wider 41.7 µm compared to 22.8 µm on a solid substrate. Hence, simulation confirms that the non-flatness of the metasurface lens affects the focal performance of the lens. To provide an intuitive explanation, the MEMS substrate is curved down in the negative z-direction which diverges the incident light, hence the focal power of the lens is effectively decreased resulting in an increase in both the focal length and focal spot size. However, we would like to point out that the effect of non-flatness of the substrate depends heavily on the substrate curvature. For example, if we were to consider the curvature in Fig. S7d, the effect of a sloped surface is like introducing a small angle deviation to the incident beam, which will be smaller compared to the effect of convex-like curvature as in Fig. S7c. The analysis for the effect of changing the incident angle can be found in [2].

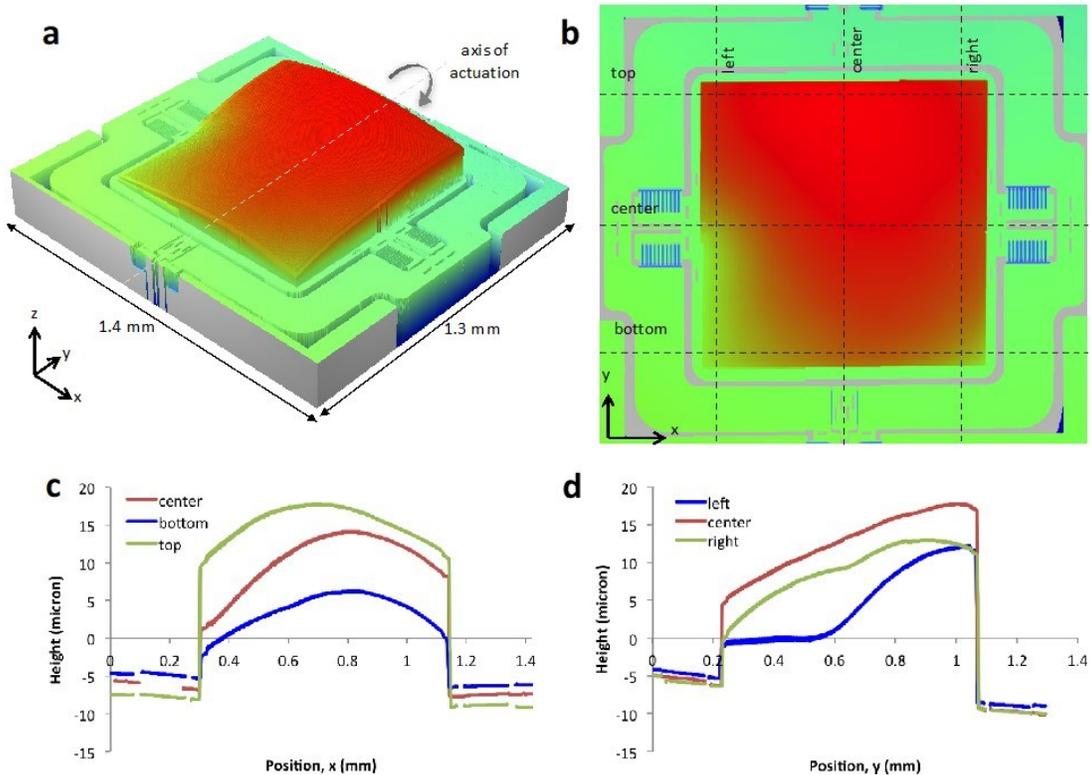

*Figure S6. (a) 3D topography of the integrated lens device measured with an optical profilometer. The green surrounding region indicates the flat peripheral portions of the MEMS frame. Red indicates higher regions. (b) Top view of the device topography. (c) and (d) shows the height distribution measured along the dashed lines marked in (b) along the x and y direction respectively.*